\documentclass[aps,floats,twocolumn,showpacs]{revtex4-1}
\usepackage[cp1251]{inputenc}
\usepackage[english]{babel}
\usepackage{amssymb}
\usepackage{amsmath}
\usepackage[pdftex]{graphicx}
\usepackage{cmap}
\usepackage[pdftex]{hyperref}
 \def\bc{\begin{center}}          \def\ec{\end{center}}
 \allowdisplaybreaks

\begin{document}
 \title{ Simulations of electromagnetic emissions produced in a thin plasma by a continuously injected electron beam
}
 \author{V.V.Annenkov, I.V.Timofeev, E.P.Volchok}
 \affiliation{Budker Institute of Nuclear Physics SB RAS, 630090, Novosibirsk, Russia \\
 Novosibirsk State University, 630090, Novosibirsk, Russia}
 \begin{abstract}

In this paper, electromagnetic emissions produced in a thin beam-plasma system are studied using two-dimensional particle-in-cell simulations. For the first time, the problem of emission generation in such a system is considered  in the realistic formulation allowing for the continuous injection of a relativistic electron beam through the plasma boundary. Specific attention is given to the thin plasma case in which the transverse plasma size is comparable to the typical wavelength of beam-driven oscillations. Such a case is often implemented in laboratory beam-plasma experiments and has a number of peculiarities. Emission from a thin plasma does not require intermediate generation of electromagnetic plasma eigenmodes, as in the infinite case, and is more similar to the regular antenna radiation. In this work, we determine how efficiently the fundamental and second harmonic emissions can be generated in previously modulated and initially  homogeneous plasmas.

 \end{abstract}
 \maketitle

\section{Introduction}

The problem of collective electron beam-plasma interaction plays the key role in explaining various physical phenomena in both space and laboratory plasmas. Type II and type III solar radio bursts, generation of coherent microwave and terahertz radiation, excitation of small-scale plasma turbulence in mirror traps are still the subjects of active research. Because of the abundance and complexity of the possible beam-plasma interaction scenarios, the numerical simulations are crucial to these studies. However, even with the modern computational resources at hand, the numerical models being used are usually limited to the idealized temporal problem of beam relaxation in which periodic boundary conditions for both particles and fields along the beam propagation axis are imposed. In such a problem, the beam has a fixed energy content and is thus capable of pumping plasma oscillations for a limited period of time before being trapped by the most unstable resonant waves. The subsequent evolution of the quasi-stationary nonlinear BGK-waves takes place in the absence of realistic beam pumping.

In the real problem of continuous injection of the fresh electron beam through the plasma boundary, the nonlinear stage of beam-plasma interaction differs substantially from this scenario. Since the development of beam-plasma instability in such a problem requires larger plasma lengths, numerical experiments on the beam injection have been previously limited to either 1D Vlasov \cite{sig,ume1,ume2} or 2D particle-in-cell (PIC) simulations \cite{mus,gol,man,ume3,dia}. The more complex case of low-density relativistic beam has been studied only in 1D even in the PIC approach \cite{tim}. A distinctive feature of these simulations is energy accumulation of plasma oscillations near the injector and formation of a coherent wave packet which  regularly interacts with the beam even in the developed plasma turbulence. For the low energy saturation levels of this packet, which are typical to the cosmic beams, the main role is played by the weakly turbulent processes \cite{ume2}. In the case of high-current relativistic beams, the wave energy in such a packet can significantly exceed the thermal plasma energy causing the strongly nonlinear regime of turbulence excitation \cite{tim}.  

In the present paper, we study the steady-state injection of relativistic electron beam into the magnetized plasma with the electromagnetic 2D3V PIC code. We will mainly focus on generation of electromagnetic (EM) waves in the spatially bounded plasmas typical to beam-plasma experiments in mirror traps  \cite{pos,thu,arz}. It has been experimentally found that reducing in plasma thickness down to the radiation wavelength is accompanied by the substantial increase in the radiation efficiency \cite{bur}. Recent PIC simulations \cite{tim2} of the simplified temporal problem have shown that such a thin system with a longitudinal density modulation can radiate as a plasma antenna. Indeed, in a periodically perturbed plasma, the most unstable beam-driven wave generates superluminal satellites which are able to interact resonantly with the vacuum electromagnetic waves at the skin depth. The radiation generated in such a scheme is concentrated near the plasma frequency. Since, in a thin plasma, there are no any limitations on the EM wave generation in frontal couplings of potential plasma waves, the nonlinear interaction of the beam mode with its long-wavelength satellites can result in emission near the doubled plasma frequency. It is interesting to note that similar emissions from bounded plasmas have been observed in early PIC simulations \cite{pri}, but the theoretical interpretation of the results has been given in terms of infinite plasma.

Here,  two problems of interest will be studied. In the first case, we will investigate what fraction of total beam power can be converted to the radiation power if the beam is injected into the rippled density plasma (Sec. \ref{s3}). In the second case, we will allow density perturbations to evolve self-consistently from the initially uniform state and estimate the role of antenna mechanism in such a regime (Sec. \ref{s4}). The first problem can be interesting for designing a powerful terahertz radiation source, the second one is crucial for interpreting the experimental results from the GOL-3 facility \cite{arz}.

\section{Simulation model}\label{s2}

To simulate beam-plasma interactions we use the parallel electromagnetic 2D3V particle-in-cell (PIC) code with the standard FDTD algorithm for EM fields and the leap-frog scheme for macroparticles. We also use the charge conserving Density Decomposition
scheme\cite{Esirkepov2001} to calculate the currents. The grid and time steps in our simulations are equal to  $h{=}0.04c/\omega_p$ and $\tau{=}0.02\omega_p^{-1}$, where $\omega_p{=}\sqrt{4\pi e^2 n_0/m_e}$ is the plasma frequency, $n_0$ is the unperturbed plasma density, $c$ is the speed of light and $e$ and $m_e$ are the charge and mass of electrons.

\begin{figure}[htb]
    \includegraphics[width=220bp]{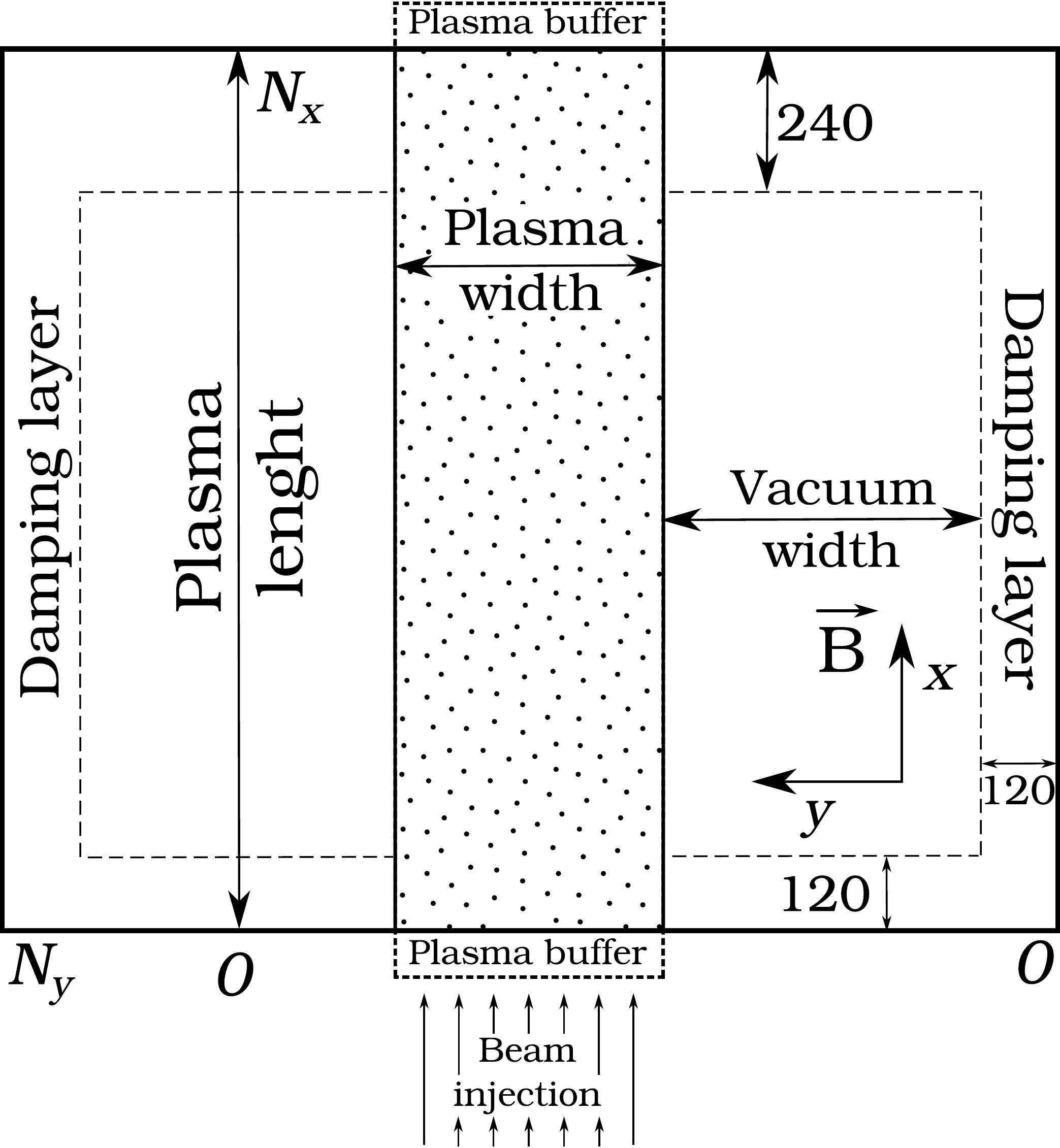} \caption{Schematic
illustration of the simulation box.}\label{Sys}
\end{figure}

The scheme of our simulation box is shown in Fig.\ref{Sys}. 
The central part of this box is initially occupied by the
plasma in which electrons and ions have arbitrary nonuniform densities $n_e(x)=n_i(x)$ and Maxwellian momentum distributions $f_{e,i}\propto \exp(-{\bf p}^2/(2 \Delta p_{e,i}^2))$ with corresponding temperatures $T_{e}= \Delta p_e^2/(2 m_e)=50$ eV and
$T_{i}{=}0$ eV. To construct the system with open boundaries in the $x$ direction, we use special buffers in which initial distributions of plasma particles are continuously supported.  In the bottom buffer ($x{<}0$), we create an additional population of beam electrons which enter the system through the one boundary and leave it through the other. In all our simulations the beam is characterized by the mean velocity $v_b=0.9c$, the relative density
$n_b/n_0=0.02$ and the Maxwellian temperature $T_{b}{=}2.56$ keV. Buffer particles can enter the system due to their thermal speeds or under the influence of the electric field $E_x$ penetrating into the buffer, whereas main particles are excluded from the simulation if they exit from the box.

The plasma column is separated from the
boundaries by vacuum regions with the characteristic width
$464 h$. Near boundaries of the simulation box we construct
damping layers in which electromagnetic fields are artificially
absorbed. Thus, our
simulation region has the characteristic sizes $L_x\times
L_y=2100h\times1328h$ and contains approximately $17\cdot10^6$
macroparticles of each plasma and beam species. The system is immersed in the uniform magnetic field
$\mathbf{B}=(B_0,0,0)$ which is determined by the electron cyclotron
frequency $\Omega_e= 0.6\omega_p$.

Further, all physical quantities are given in the dimensionless form: all frequencies are measured in $\omega_p$, wavenumbers in $c/\omega_p$, velocities in $c$, fields in $e m_e c/\omega_p$, number densities in $n_0$.

\section{Generation of EM waves in modulated plasmas}\label{s3}

Let us first consider electromagnetic radiations produced by the injection of an electron beam into the magnetized plasma with the previously modulated density $n_i=n_0+\delta n \cos (qz)$. Recent PIC simulations with periodic boundary conditions \cite{tim2,tim3} showed that, if the plasma thickness is comparable with the radiation wavelength, the power of electromagnetic emission near the frequency of the most unstable beam-driven mode can constitute the significant fraction (about 7\%) of total beam power.  It was also shown that the main features of observed radiation can be explained by the mechanism of plasma antenna. In the present study, by modelling the open beam-plasma system we  will examine how adequately the proposed theory can describe the fundamental electromagnetic emission in the realistic problem of steady-state beam injection through the plasma boundary.  In addition, we will determine the efficiency of nonlinear processes responsible for the generation of EM waves at the doubled plasma frequency.

\subsection{Fundamental emission}

If in the plane plasma layer with the thickness $2l$ the beam excites the longitudinal wave with the frequency $\omega_b$ and wavenumber $k_{\|}=\omega_b/v_b$,  scattering of this wave on the density perturbation with the number $q$ produces forced plasma oscillations with  $(\omega_b, k_{\|}-q)$. Such oscillations can have superluminal phase velocities and are able to get in resonance with vacuum electromagnetic waves running obliquely to the plasma layer. Such a resonant interaction becomes possible only if the period of plasma density modulation does not much differ from the wavelength of the beam-driven wave ($1-v_b<q/k_{\|}<1+v_b$). The ratio between $q$ and $k_{\|}$ inside this range fixes the unique radiation angle
\begin{equation}
	\theta=\arctan\left(\frac{\sqrt{v_b^2-(1-q/k_{\|})^2}}{1-q/k_{\|}}\right).
\end{equation}
In the particular case $q=k_{\|}$, the radiation emerges from the plasma in the purely transverse direction and is polarized along the ambient magnetic field (${\bf E}||{\bf B}_0$). Since in the hydrodynamic regime of beam-plasma instability the frequency of the dominant wave $\omega_b$ is less than the plasma frequency,
\begin{equation}
	\omega_b=1-\frac{n_b^{1/3}}{2^{4/3} \gamma_b},
\end{equation}
the produced radiation can effectively interact with plasma currents only at the skin-depth ($\gamma_b$ is the beam relativistic factor). The total power of EM emission in the realistic problem of beam injection can be obtained by generalizing the result of the work \cite{tim3} to the case of the arbitrary nonuniform amplitude of the dominant beam-driven mode $E_0(x)$: 
\begin{equation}\label{p}
	\frac{P_{rad}}{P_b}=\frac{\delta n^2 F(l)}{8 (\gamma_b-1)n_b v_b \sqrt{1-\omega_b^2}} \int\limits_0^{L} E_0^2(x) dx,
\end{equation}
where the factor
\begin{equation}
	F(l)=\frac{\sinh^2(\varkappa l)}{ \varkappa l \left[\omega_b^2+\sinh^2(\varkappa l) \right]}
\end{equation}
describes how this power depends on the plasma half-width $l$,
$\varkappa=\sqrt{1-\omega_b^2}$ defines the reciprocal skin depth, $P_b$ --- the power of injected beam, and $L$ --- the length of radiating plasma region (here, all number densities are measured in units of $n_0$, velocities  in the speed of light $c$, frequencies in $\omega_p$, scale lengths in $c/\omega_p$, fields in $m_e c \omega_p/e$). The paper \cite{tim3} deals with the spatially uniform case ($E_0=const$), that is why applying its results to the system with open boundaries means  the rude approximation for the integral (\ref{p}) ($\int \sim E_0^2 L$). In the previously proposed model, it was also assumed that the amplitude of the dominant unstable wave is saturated at the level of beam trapping $E_0\sim \gamma_b^3 \Gamma^2 v_b$, where $\Gamma$ is the growth rate of the two-stream instability, and the radiating zone is limited by the typical length of the coherent wave packet $L\sim 3v_b/\Gamma$ in which this trapping occurs. In this study, we will calculate the radiation efficiency using the realistic profiles of wave energy, which are self-consistantly formed during the beam injection into the plasma layer. 

To check theoretical predictions let us perform the numerical experiment in which the beam is continuously injected into the plasma with the thickness $2l=6.4$ and the length  $L=83.2$. The plasma has the longitudinal density modulation  with the period $\lambda_q=2 \pi v_b/\omega_b\approx 5.93$ and the depth $\delta n=0.2$. 
\begin{figure}[htb]
\bc\includegraphics[width=220bp]{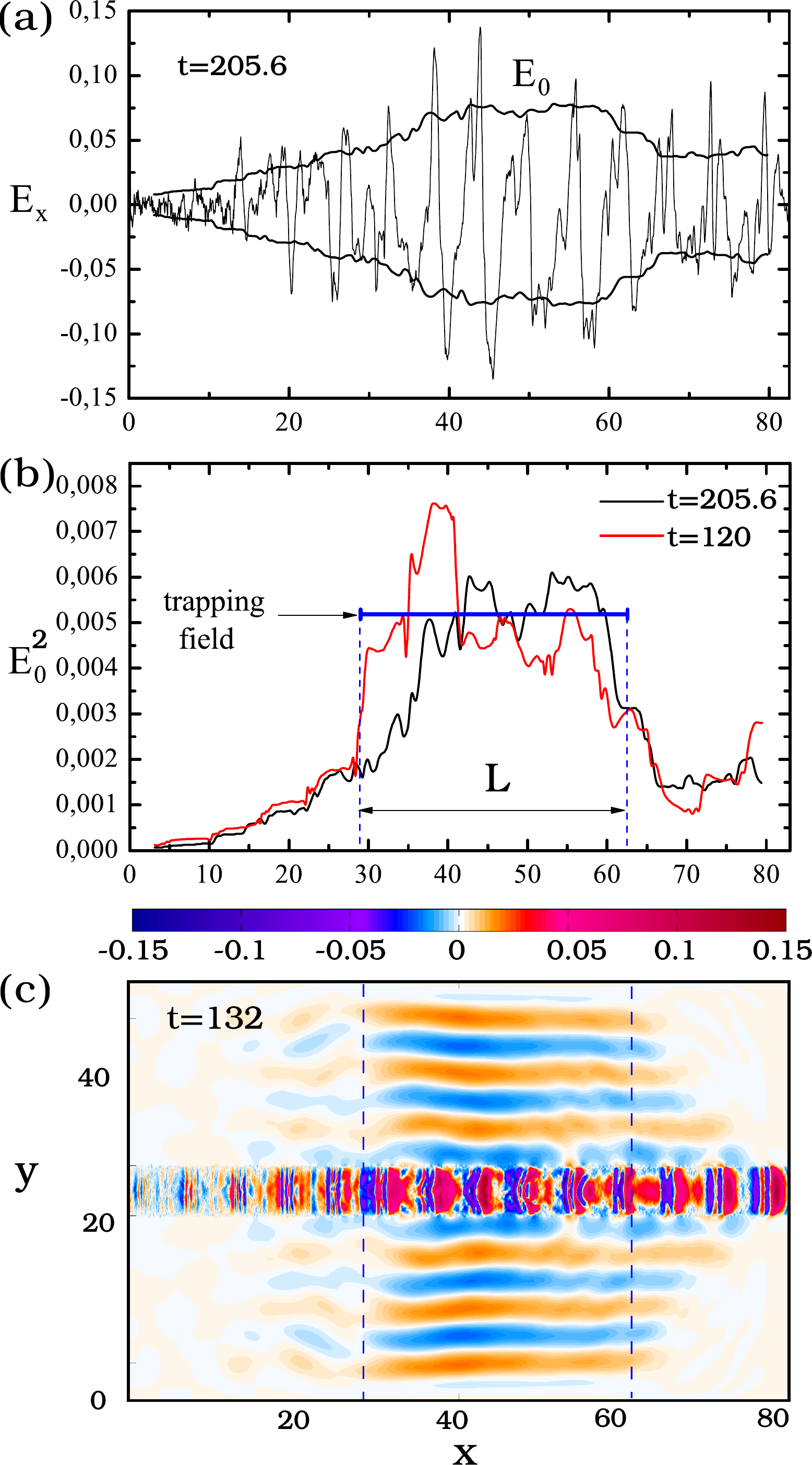} \ec \caption{(a) The averaged amplitude $E_0(x)$ of the beam-driven mode and the field $E_x(x)$ in the center of plasma layer. (b) Profiles of resonant wave energy $E_0^2(x)$ in different moments of time. (c) The electric field map $E_x(x,y)$.}\label{fig1}
\end{figure}
Let us first clarify the accuracy of theoretical estimates for the typical length and the saturation energy of the wave packet in which the beam is trapped. It is seen from Fig. \ref{fig1}a that the amplitude of the beam-driven wave $E_0$, averaged in the longitudinal direction over the length $\lambda=2 \pi v_b/\omega_b$ and across the plasma over the width  $2 l$, slightly differs from the envelope of the maximal field $E_x$ measured in the center of plasma layer. It means that the spectrum of unstable plasma oscillations is actually dominated by the longitudinal wave, and its amplitude slightly changes across the plasma. More clearly the localization of wave energy is shown in Fig. \ref{fig1}b. Here, for instance, we demonstrate profiles $E_0^2(x)$ in different moments of time. It is seen that the length of appearing wave packet is well agreed with the estimate $L=3v_b/\Gamma$, and the energy of resonant waves is really saturated at the level $E_0^2=\left(\gamma_b^3 \Gamma^2 v_b\right)^2$ determined by the beam trapping effect. In theory, the square of the amplitude governs the radiation intensity. The map of electric field $E_x(x,y)$ presented in Fig. \ref{fig1}c shows that the region of the most intense electromagnetic emission does really coincide with the region of plasma wave energy localization. By integrating the numerical profiles  $E_0^2(x)$ and substituting these results in (\ref{p}), we can theoretically predict what part of the total beam power in some given moment of time is converted into the emission power of EM waves. The temporal dependence of such a relative radiation power  $P_{rad}/P_b$ is shown in Fig. \ref{fig2}.
\begin{figure}[htb]
\bc\includegraphics[width=240bp]{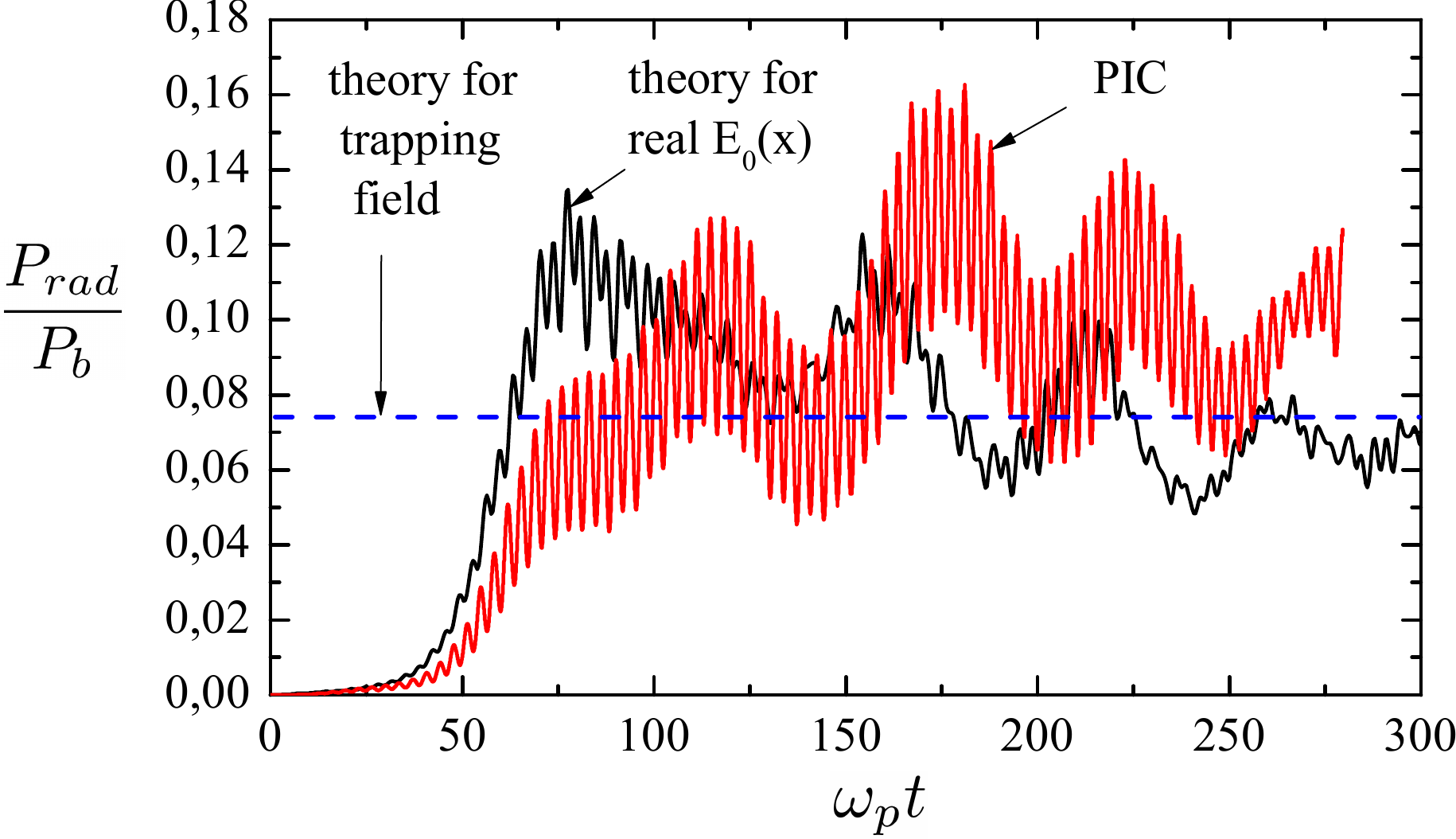} \ec \caption{The radiation efficiency $P_{rad}/P_b$ as a function of time in PIC simulations (red line), in theory for real profiles $E_0(x)$ (black line), and in theory for uniform $E_0$ (blue dashed line).}\label{fig2}
\end{figure}
One can see that the high radiation efficiency ($\sim 10 \%$) predicted theoretically is also confirmed by numerical simulations. The weak dependence of wave amplitude on the transverse coordinate that is not taken into account in our theory results in generation of EM waves with the transverse polarization (${\bf E}\bot {\bf B}$). The contribution of this radiation is the reason why the observed emission power exceeds the theoretical value for the O-mode. 

Thus, injection of an electron beam in a thin modulated plasma can result in highly efficient generation of narrowly directed EM emission with the intensity localized on the length $L\approx 3v_b/\Gamma$ and the frequency tied to the frequency of the most unstable beam-driven mode ($\omega_b\approx \omega_p$). Since the typical power of high-current electron beams can reach tens of gigawatts, such a scheme is able to generate the sub-terahertz radiation of the gigawatt level  during the ions time-scale. 
\begin{figure*}[htb]
\bc\includegraphics[width=500bp]{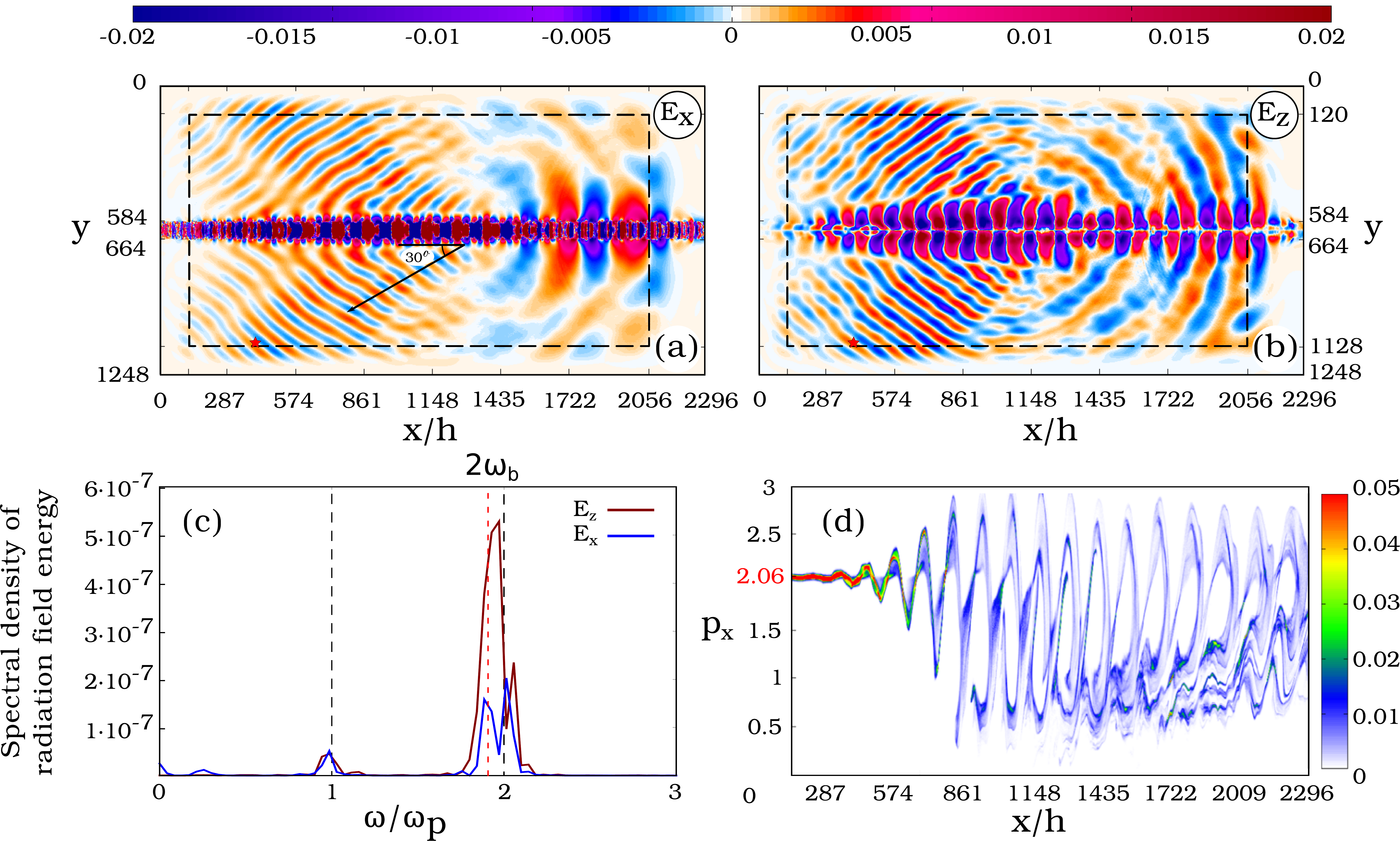} \ec \caption{Generation of $2 \omega_b$-radiation in the modulated plasma. (a),(b) -- The maps of electric fields $E_x$ and $E_z$. (c) -- The frequency spectrum of $E_x$ measured in the single point lying on the damping layer (red star). (d) -- The phase space ($x,p_x$) of beam electrons.}\label{fig3}
\end{figure*}

\subsection{Second harmonic emission}

The second harmonic emission can be produced by the nonlinear coupling of the dominant beam-driven mode $(\omega_b, k_{\|})$ with its long-wavelength satellite ($\omega_b, k_{\|}-q$) arising from scattering on the periodically perturbed ion density:
\begin{equation}
	(\omega_b, k_{\|})+(\omega_b, k_{\|}-q)\rightarrow (2\omega_b, 2k_{\|}-q).
\end{equation}
According to the plasma antenna mechanism, the resulting wave of electric current $j\propto \exp(i(2k_{\|}-q)x-i2\omega_b t)$ can be the source for electromagnetic emissions only in the limited range of $q$ ($1-v_b<q/2k_{\|}<1+v_b$). This implies that, for the modulation period corresponding to the range
\begin{equation}\label{d}
1+v_b<q/k_{\|}<2(1+v_b), 
\end{equation}
most of radiation escaping from the plasma should be concentrated near the second harmonic of the pump wave. The angle of this radiation is given by the value
\begin{equation}\label{th}
	\theta=\arctan\left(\frac{\sqrt{v_b^2-(1-q/2k_{\|})^2}}{1-q/2k_{\|}}\right).
\end{equation}

In order to observe the enhancement of the second harmonic emission let us perform the simulation experiment with the same beam injected in the plasma layer with the width $2l=3.2$. The period of plasma density modulation $\lambda_q=1.67$ corresponds to the wavenumber $q$ lying in the range (\ref{d}), and the depth is chosen at the former level $\delta n=0.2$. In this case, EM waves should be radiated at the angle $\theta\approx 30^0$ and run towards the injected beam. From the maps of electric fields $E_x$ (Fig. \ref{fig3}a) and $E_z$ (Fig.\ref{fig3}b) one can see that, in the region of intense beam-plasma interaction, the narrow beam of EM waves is really radiated, but the angle of this emission appears to be higher ($\theta\approx 40^0-50^0$).
The point is that our simplified formulas loses its accuracy in the regime of very strong beam-plasma instability when the beam at the trapping stage has the significant change in velocity. The beam phase-space $(x,p_x)$ (Fig. \ref{fig3}d) demonstrates how beam electrons mix under the field of high-amplitude resonant wave. It is seen that, inside the radiating plasma region,   the beam velocity distribution occupies the range from $v_b=0.5$ to $v_b=0.94$. If we assume that the wavenumber $k_{\|}=\omega_b/v_b$ of the nonlinearly saturated wave is determined by some mean velocity $v_b\approx 0.75-0.8$, the formula (\ref{th}) reproduces the observed values of  $\theta$.

\begin{figure}[htb]
\bc\includegraphics[width=230bp]{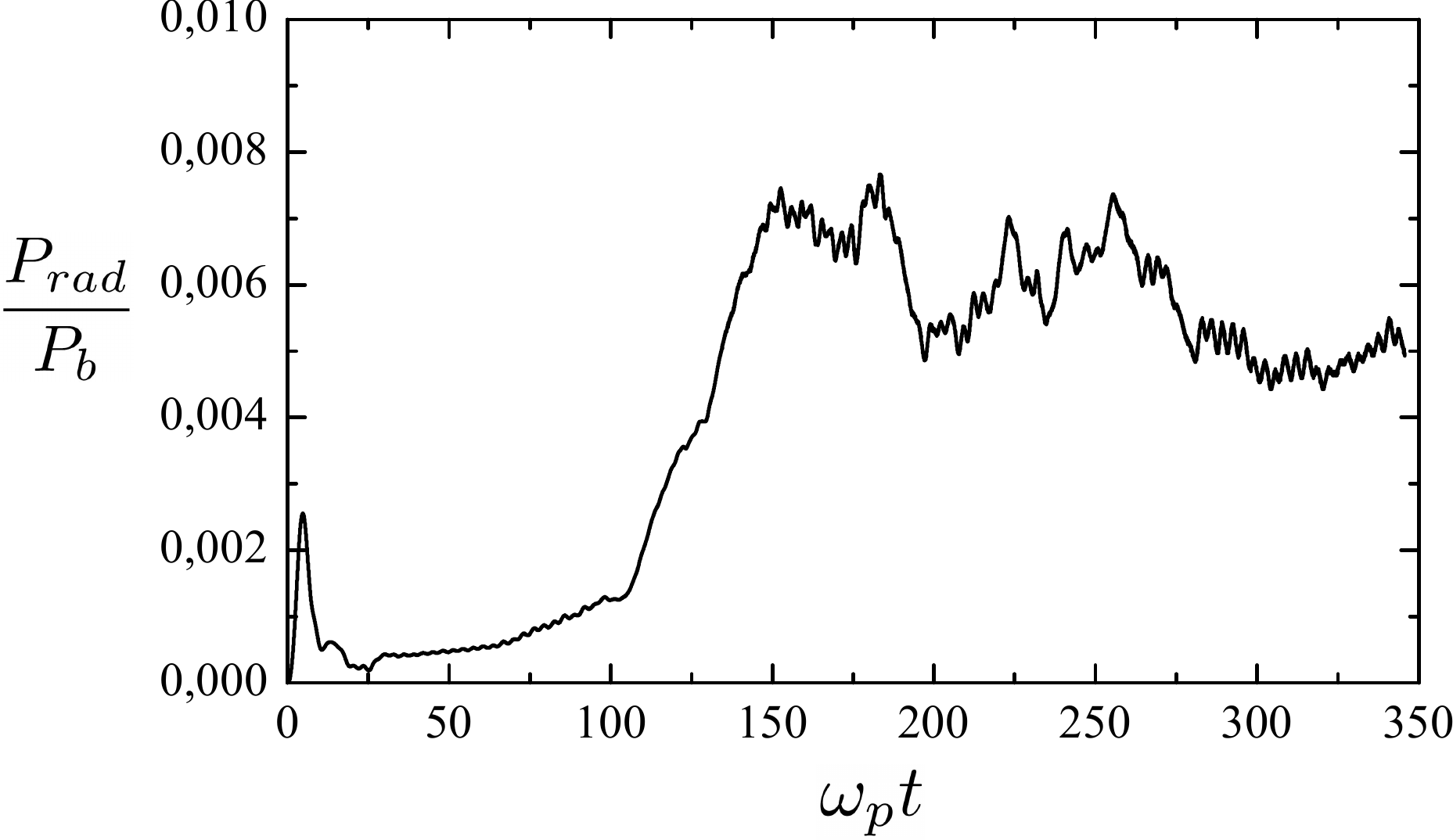} \ec \caption{The generation efficiency for the $2\omega_b$-radiation as a function of time.}\label{2w}
\end{figure}

Measuring the histories of emission fields in the certain point near the damping layer (its position is indicated by the red star in Fig. \ref{fig3}a), we can find the frequency spectrum of generated EM waves (Fig. \ref{fig3}c). It is seen that the frequency lies near $2\omega_b$, and most of intensity is concentrated in X-polarized waves with the field $E_z$. Measuring the energy lost by EM waves in transverse and input damping layers, we obtain the relative power of second harmonic emission (Fig. \ref{2w}). Figure shows that generation of such emission reaches 0.7\%. 

\section{EM emission from initially iniform plasma}\label{s4}

Let us now study radiation processes under the beam injection into the plasma with the initially uniform density. In this case, the turbulent plasma is characterized by the wide spectrum of ion density fluctuations, but antenna radiation can be generated by those of them, scattering on which creates superluminal waves of electric current. Fluctuations with wavenumbers $q\in[k_{\|}-\omega_b; k_{\|}+\omega_b]$ are responsible for the fundamental emission, whereas the range  $q\in[2(k_{\|}-\omega_b); 2(k_{\|}+\omega_b)]$ produces the emission near the doubled plasma frequency.
\begin{figure*}[htb]
\bc\includegraphics[width=500bp]{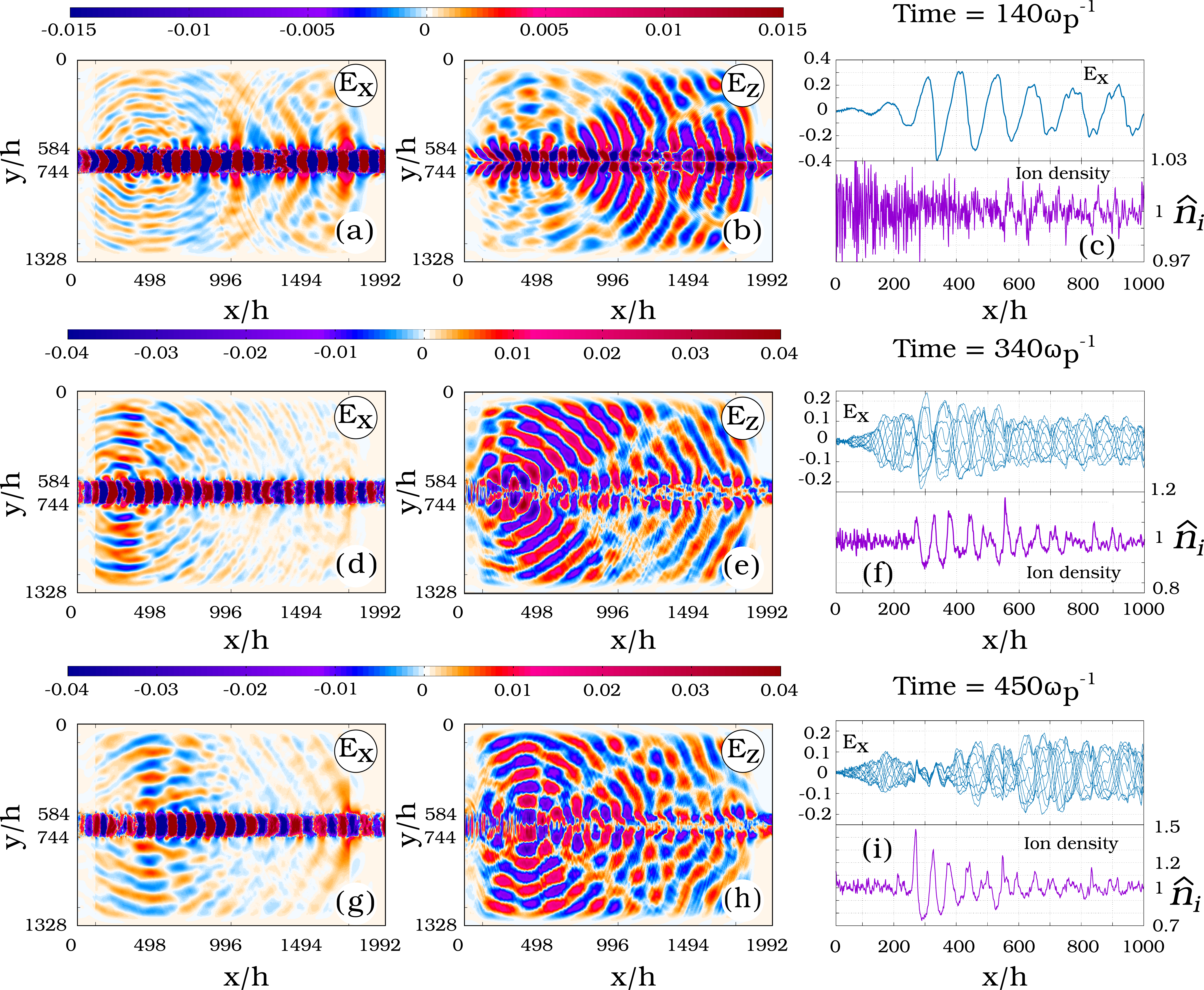} \ec \caption{The case of injection into the uniform plasma. (a),(d),(g) -- The maps of electric field $E_x$ in different moments of time. (b),(e),(h) -- The maps of electric field $E_z$. (c),(f),(i) -- Electric fields $E_x$ in the center of plasma layer and corresponding $y$-averaged density profiles in different moments of time. }\label{rad2}
\end{figure*}

\begin{figure}[htb]
\bc\includegraphics[width=230bp]{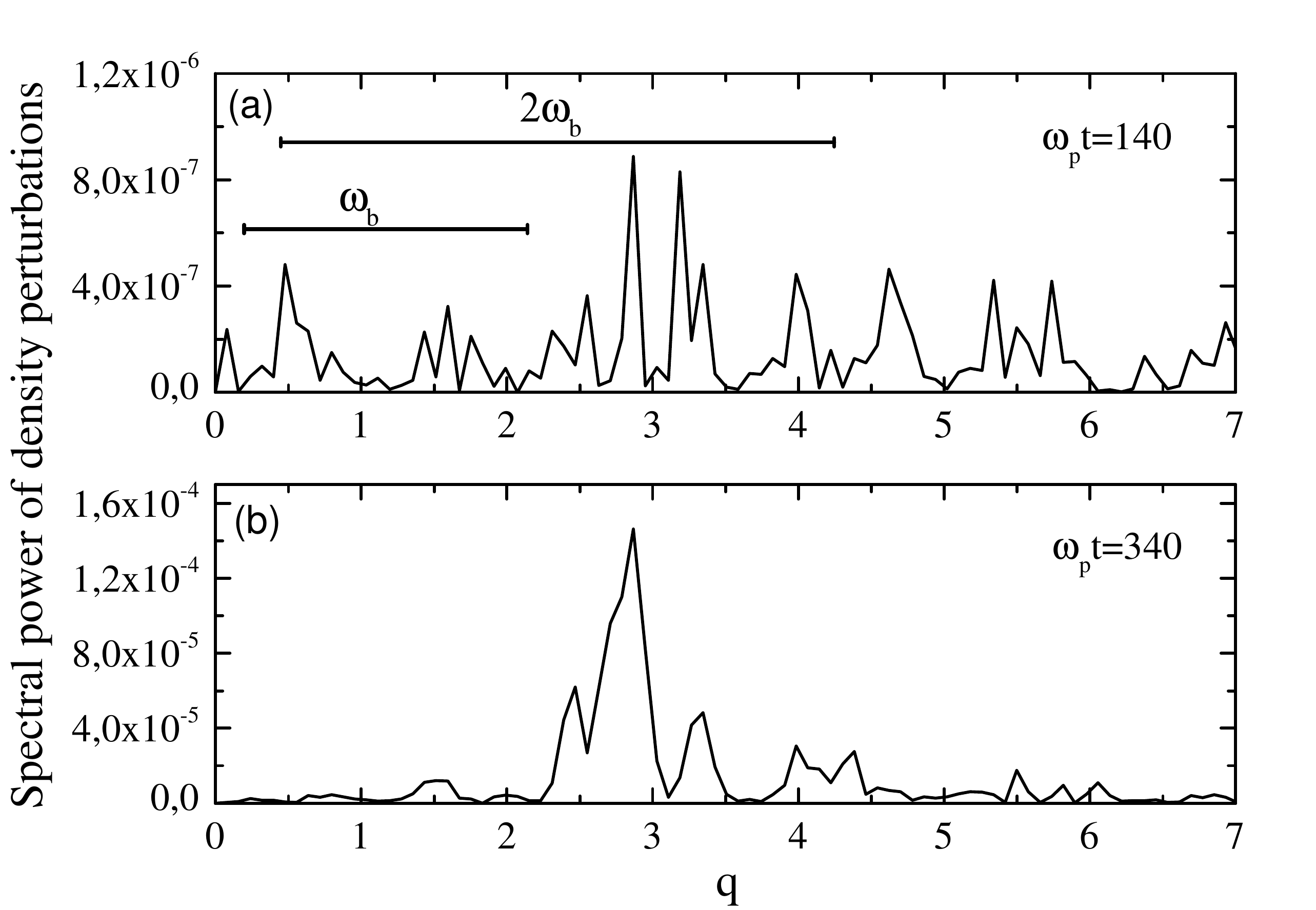} \ec \caption{The spectrum of density perturbations in various moments of time.}\label{sp}
\end{figure}

Let us follow the dynamics of emissions produced by the beam injection into the uniform plasma layer with the thickness $2l=6.4$. The build-up of the beam-plasma instability results in formation of the quasi-one-dimensional wave packet in which the wave energy is nonlinearly saturated by beam trapping. The amplitude of travelling resonant wave in this packet grows up to the high value, at which the trapping zone extends down to the zero speed. Under the field of such an energetic wave, the modulational instability is developed. By the moment $\omega_p t=140$, the spectrum of $y$-averaged density fluctuations is dominated by the modulational perturbation with the wavenumber   $q\approx 2.85$ (Fig. \ref{sp}). 
Since the local wavenumber of the beam-driven mode inside the region ($x/h_x \in[300,550]$) reaches the value $k_{\|}\approx 1.4$, the growing density modulation can stimulate only  the $2\omega_b$-radiation. This transversely propagating radiation is really observed in the map of electric field $E_x$ (Fig. \ref{rad2}a). In the remote region ($x/h_x \in[550,1100]$), the local $k_{\|}$ decreases ($k_{\|}\approx 1.17$).  In this case, the $\omega_b$-radiation is only sensitive to long-wavelength density perturbations within the range $q\in [0.2, 2.1]$ in which the dominant one ($q\approx 0.5$) is likely determined by the noise properties (Fig. \ref{sp}a). In any way, scattering of the beam mode on such a perturbation explains the radiation angle $\theta\approx 45^0$ visible in the map $E_z(x,y)$ (Fig. \ref{rad2}b).

In the moment $\omega_p t=340$, the modulational instability with $q\approx 2.85$ reaches the nonlinear stage which is accompanied by trapping of plasma oscillations in arising density wells. Fig. \ref{rad2}f shows that standing oscillations of longitudinal electric field  $E_x$  inside the nearest to the injector density cavern appear. The wavelength of these trapped oscillations is automatically adjusted to the modulation lengh, which allows the plasma antenna to radiate in the transverse direction at the frequency $\omega_b$ (Fig. \ref{rad2}d). More intensive radiation with the transverse polarization $E_z$ comes from the plasma at an acute angle (Fig. \ref{rad2}e) which is possibly explained by some phase shifting between trapped oscillations in neighboring density wells.

\begin{figure}[htb]
\bc\includegraphics[width=230bp]{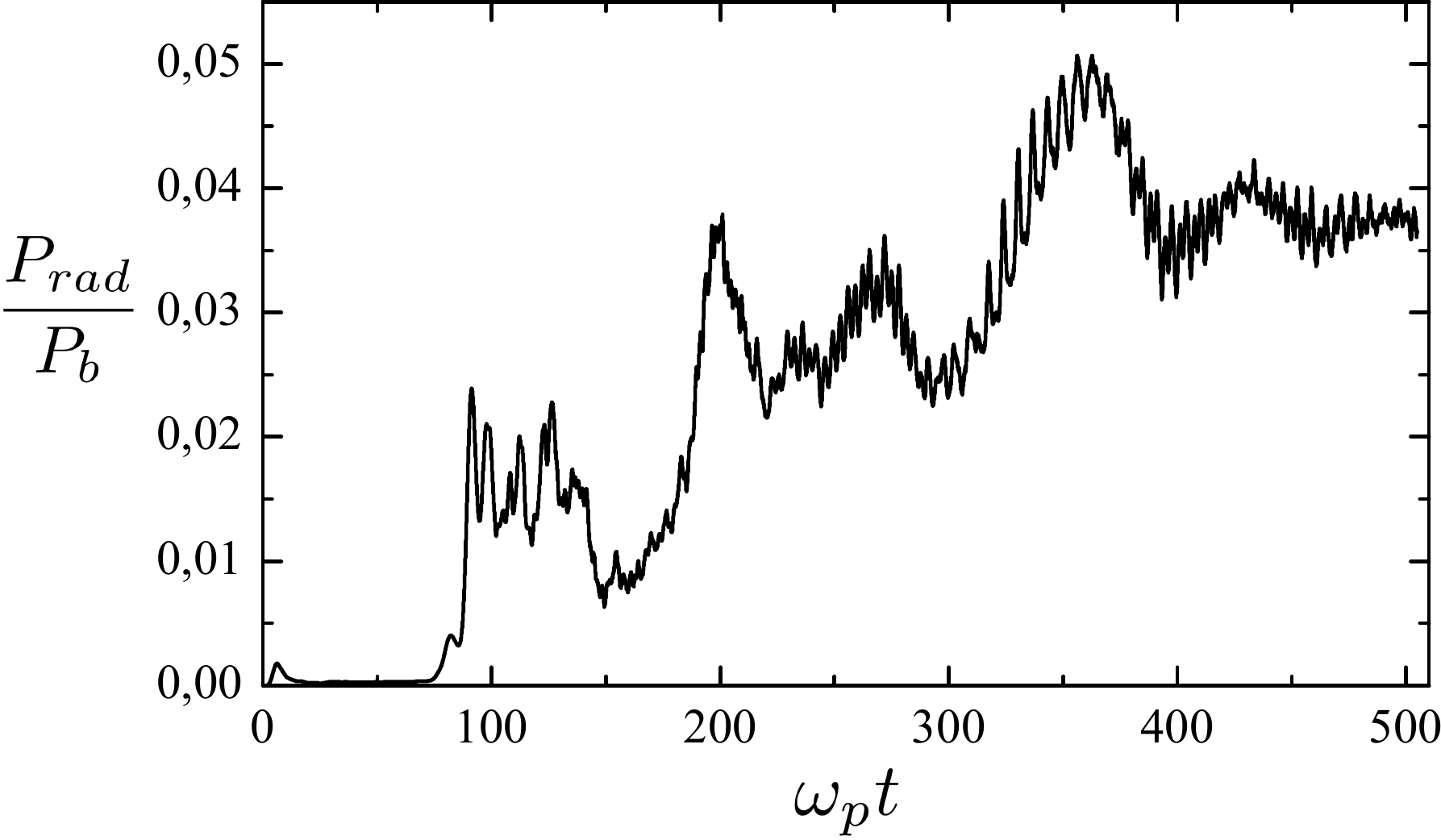} \ec \caption{The radiation efficiency in the case of self-consistent density fluctuations.}\label{hom}
\end{figure}

Since standing oscillations with $k\approx 2.8$ fall out of the resonance with the beam, they burn out with the deepening of the density well (Fig. \ref{rad2}i). The region of intense beam pumping is gradually removed from the injector to the regions with less gradients of plasma density. The powerful $\omega_b$-radiation at this stage is generated by trapped plasma oscillations and predominantly directed across the plasma  (Fig. \ref{rad2}h). It is seen from Fig. \ref{hom} that generation of EM waves becomes more efficient when  the nonlinear stage of modulational instability results in trapping of beam-driven oscillations. Upon reaching this stage, the conversion of the beam power to the emission power is established at the level 4\%.  

\section{Summary}

For the first time, generation of electromagnetic waves under the steady-state injection of relativistic electron beam into the magnetized finite-size plasma is studied using PIC simulations. It is shown that the previously proposed theoretical model based on the plasma antenna mechanism adequately describes electromagnetic emissions produced in the open beam-plasma system with the typical width comparable to the emission wavelength. Numerical experiments on the beam injection into the previously modulated plasma  show that the power of the fundamental electromagnetic emission constitutes 10\% of the total beam power, whereas the second harmonic radiation efficiency in such a scheme reaches 0.7\%. Since the power of high-current electron beams can reach tens of gigawatt, the proposed scheme seems attractive for the generation of sub-terahertz radiation with the gigawatt power level.

It is also found that the longitudinal plasma density modulation playing the key role in the antenna mechanism may appear self-consistently due to the modulational instability of the high-amplitude beam-driven mode. It is shown that, if the most unstable  density perturbation does not initially result in excitation of superluminal satellites, the later nonlinear stage of modulational instability is accompanied by trapping of plasma oscillations by individual density wells in which the condition for the efficient antenna radiation $q=k_{\|}$ is automatically fulfilled. In such a regime, the transversely propagating radiation is predominantly polarized across the magnetic field,  concentrated near the plasma frequency and reaches the efficiency  4\%. The antenna mechanism should play an important role in the laboratory beam-plasma experiments at the GOL-3 mirror trap, in which,  according to some estimates, the close level of radiation efficiency has been observed.

\begin{acknowledgments}
This work is supported by RFBR (grant 15-32-20432).  
\end{acknowledgments}

\end{document}